\documentclass[conference]{IEEEtran}

\usepackage{xcolor}

\usepackage{amsmath,amsfonts}
\usepackage{amssymb}

\usepackage{siunitx}

\definecolor{olivegreen}{rgb}{0, 0.6, 0}
\definecolor{redorange}{HTML}{FF5349}
\definecolor{blue(ncs)}{rgb}{0.0, 0.53, 0.74}
\definecolor{navy}{HTML}{273BE2}

\definecolor{black}{HTML}{000000}
\definecolor{white}{HTML}{ffffff}
\definecolor{color1}{HTML}{ACE5EE}
\definecolor{color2}{HTML}{0093AF}
\definecolor{color3}{HTML}{CC0000}
\definecolor{color4}{HTML}{0087BD}
\definecolor{color5}{HTML}{333399}
\definecolor{color6}{HTML}{20B2AA}

\usepackage{xspace}

\usepackage{graphicx}
\usepackage[export]{adjustbox}
\usepackage{wrapfig}

\usepackage{enumitem} 
\usepackage{tcolorbox}

\usepackage{verbatim}

\usepackage{booktabs}
\usepackage{multicol}
\usepackage{multirow}
\usepackage{makecell}

\usepackage{nicefrac}

\usepackage{algorithm}
\usepackage{algorithmic}

\usepackage[hidelinks]{hyperref}

\usepackage[noabbrev, capitalize]{cleveref} 

\usepackage{pifont}

\usepackage{textcomp}

\usepackage[a4paper, total={184mm,239mm}]{geometry}
\def\BibTeX{{\rm B\kern-.05em{\sc i\kern-.025em b}\kern-.08em
    T\kern-.1667em\lower.7ex\hbox{E}\kern-.125emX}}

\usepackage[sorting=none, maxcitenames=1,mincitenames=1,uniquelist=false, style=ieee]{biblatex}

\AtBeginBibliography{\footnotesize}
\renewbibmacro{in:}{}
\newcommand{\thiswork}{Pipette\xspace}

\newcommand{\latencyestimate}{latency estimator\xspace}

\newcommand{\LatencyEstimate}{Latency Estimator\xspace}

\newcommand{\autoconfig}{fine-grained worker dedication\xspace}
\newcommand{\AutoConfig}{Fine-grained Worker Dedication\xspace}

\newcommand{\memestimate}{memory estimator\xspace}
\newcommand{\Memestimate}{Memory estimator\xspace}
\newcommand{\MemEstimate}{Memory Estimator\xspace}

\definecolor{myblue}{RGB}{3, 110, 184}
\definecolor{mygreen}{RGB}{49, 115, 58}
\definecolor{myorange}{RGB}{243, 151, 0}
\definecolor{mypurple}{RGB}{143, 77, 158}

\begin{document}

\title{\thiswork: Automatic Fine-grained Large Language Model Training Configurator for Real-World Clusters}

\author{%
\IEEEauthorblockN{%
Jinkyu Yim\IEEEauthorrefmark{2}\textsuperscript{,1},
Jaeyong Song\IEEEauthorrefmark{2}\textsuperscript{,1},
Yerim Choi\IEEEauthorrefmark{4},
Jaebeen Lee\IEEEauthorrefmark{4},
Jaewon Jung\IEEEauthorrefmark{2},
Hongsun Jang\IEEEauthorrefmark{2}
and Jinho Lee\IEEEauthorrefmark{2}\textsuperscript{,*}}\vspace{2pt}%
\IEEEauthorblockA{%
\IEEEauthorrefmark{2}\textit{Department of Electrical and Computer Engineering, Seoul National University}\\%
\IEEEauthorrefmark{4}\textit{Samsung Electronics}}
\vspace{2pt}%
\IEEEauthorblockA{%
\{skyson00, jaeyong.song\}@snu.ac.kr, \{yr9.choi, jbeen.lee\}@samsung.com, \{hongsun.jang, jungjaewon, leejinho\}@snu.ac.kr}}
\maketitle

\begin{abstract}
Training large language models (LLMs) is known to be challenging because of the huge computational and memory capacity requirements.
To address these issues, it is common to use a cluster of GPUs with 3D parallelism, which splits a model along the data batch, pipeline stage, and intra-layer tensor dimensions.
However, the use of 3D parallelism produces the additional challenge of finding the optimal number of ways on each dimension and mapping the split models onto the GPUs.
Several previous studies have attempted to automatically find the optimal configuration, but 
many of these lacked several important aspects.
For instance, the heterogeneous nature of the interconnect speeds is often ignored.
While the peak bandwidths for the interconnects are usually made equal, the actual attained bandwidth varies per link in real-world clusters.
Combined with the critical path modeling that does not properly consider the communication, they easily fall into sub-optimal configurations.
In addition, they often fail to consider the memory requirement per GPU, often recommending solutions that could not be executed.
To address these challenges, we propose \emph{\thiswork}, which is an automatic fine-grained LLM training configurator for real-world clusters. 
By devising better performance models along with the memory estimator and fine-grained individual GPU assignment, 
\thiswork achieves faster configurations that satisfy the memory constraints.
We evaluated \thiswork on large clusters to show that it provides a significant speedup over the prior art.
The implementation of \thiswork is available at \textit{\url{https://github.com/yimjinkyu1/date2024_pipette}}.
\end{abstract}


\begingroup
\renewcommand\thefootnote{1}\makeatletter\def\Hy@Warning#1{}\makeatother\footnotetext{Co-first authors. \ \ \ \ \textsuperscript{*}Corresponding author.}
\endgroup

\section{Introduction}
\label{sec:intro}
\begin{figure}[t]
\centering
 \includegraphics[width=0.9\columnwidth]{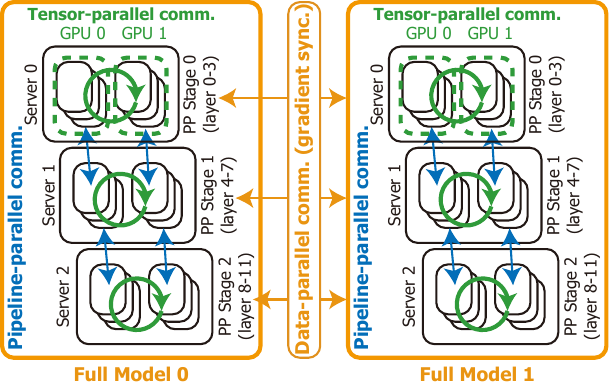}
 \caption{Example 3D parallelism configuration with 12 GPUs.
 } \vspace{-5mm}
 \label{fig:intro}
\end{figure}

It is evident that both the industry and academia are experiencing a remarkable surge in large language models (LLMs)~\cite{bert, gpt2, gpt3}.
As the model size grows, the performance (i.e., accuracy) of an LLM is known to continuously improve for various tasks.
To provide its huge computational power and memory capacity, 
researchers often use 3D parallelism, which divides a model along three dimensions (tensor-wise, layer-wise, and batch-wise) as depicted in \cref{fig:intro}.
This has been demonstrated to be efficient at utilizing a massive number of GPUs~\cite{gpt3, zero, sc21megatron}.
However, the exact configuration of the GPUs toward the optimal performance still remains to be solved.

Heuristic approaches~\cite{megatrongithub, zero} rely on manual rules for the configuration,  derived from insights gained by domain experts through numerous trials.
To reduce the effort to manually find an adequate configuration, a recent trend~\cite{piper, amp} has been to automatically search for near-optimal configurations.
Most of them profile the computation time and then integrate it into their pipeline latency models to get latencies of configurations.

Unfortunately, we diagnose that these methods tend to have three main limitations that restrict their practical use in the field.
\begin{enumerate}
\item \emph{Static Interconnect Assumption.}
The existing methods simply assume that the interconnects between the servers are static, with a fixed bandwidth and latency. 
However, the actual communication latency in a real cluster exhibits heterogeneity among the links~\cite{plink, moody2009contention, vazhkudai2018design}, which could yield unexpected straggler effects.

\item \emph{Hidden Critical Path.}
The existing methods construct latency models on the 3D parallelism but miss some critical paths.
This comes from the discrepancy between the latency model and modern scheduling.
While the state-of-the-art latency models~\cite{amp,varuna} assume outdated scheduling methods to achieve maximal throughput, the de facto standard is to use memory-efficient scheduling~\cite{pipedream, sc21megatron} to relieve the memory capacity requirements.

\item \emph{Out-of-Memory Configurations.}
The configurations recommended by the automated tools often require more memory per GPU than what is physically available. 
This is because those methods do not consider the memory usage of LLM~\cite{amp} at all or fail to estimate it~\cite{varuna}.
\end{enumerate}

To address the above limitations, we propose \emph{\thiswork}, an automatic fine-grained 3D parallelism configurator for real-world clusters.
We devise the following three novel schemes.
First, to fully take the heterogeneous interconnect latency into account, \thiswork performs \emph{\autoconfig} to determine the location of individual workers in the cluster.
Second, \thiswork uses an elaborately designed \emph{\latencyestimate} based on the refined latency model, which addresses the aforementioned hidden critical path problem of prior works.
For the communication terms in the latency model, \latencyestimate profiles the actual latency of each interconnect in a real-world cluster.
Last, to only produce configurations that meet the per-GPU memory constraint, we introduce accurate \emph{\memestimate}, based on a simple ML model.
We evaluated \thiswork on clusters with up to 128 NVIDIA V100 or A100 GPUs connected via NVLink and Infiniband, and it provided significant up to 1.46$\times$ speedup over the prior art AMP~\cite{amp}.
The implementation of \thiswork is open-sourced at \textit{\url{https://github.com/yimjinkyu1/date2024_pipette}}.

\section{Background and Related Work}
\label{sec:background}

\subsection{3D Parallelism and Pipeline Schedule}
\label{sec:background:3d}
\emph{3D parallelism}~\cite{megatron-lm, sc21megatron, zero} is a common parallelization technique used to train LLMs with tens to thousands of GPUs in a cluster.
It splits a model into several parts and distributes them to each worker (i.e., GPU) using \emph{pipeline (PP)}, \emph{tensor (TP)}, and \emph{data (DP) parallelism}.
\cref{fig:intro} depicts an example 3D parallelism configuration with a 3-way pipeline, 2-way tensor, and 2-way data parallelism.
Because the entire model comprises tens to hundreds of layers, pipeline parallelism divides the layers into several stages.
The peer-to-peer pipeline parallel communications ({\color{myblue}blue arrows}) fulfill the dependency among the pipeline stages.
Tensor parallelism further splits the layers at an intra-layer level.
It utilizes all-reduce communication ({\color{mygreen}green arrows}) inside each layer.
Because this all-reduce communication introduces a severe overhead, tensor parallel ways are usually placed within a server.
Data parallelism, a widely used technique to parallelize deep learning models, can be orthogonally applied on top of the above two parallelization methods.
It requires parameter gradient synchronization ({\color{myorange}orange arrows}), which again uses all-reduce communication.

\cref{fig:background} illustrates one training iteration of the example configuration in \cref{fig:intro} with six microbatches.
Usually, the key is to hide the communication latency by overlapping it with the computation and reducing the idle resource (i.e., bubbles) by using microbatches.
One overhead from such a schedule is the high memory capacity requirement, because each GPU has to store activations and gradients of all six microbatches for backward passes.
Therefore, recent studies~\cite{gpt3, sc21megatron, zero, optimuscc} on LLM training adopted the memory-efficient schedule (i.e., 1F1B) to reduce the memory usage.
As shown in \cref{fig:background}b, interleaving the forward and backward one by one dramatically reduces the memory capacity requirements because the microbatch data can be dropped from the memory after performing backpropagation.

\subsection{Automatic Configuration and Latency Model}
\begin{figure}[t]
\centering
    \includegraphics[width=.95\columnwidth]{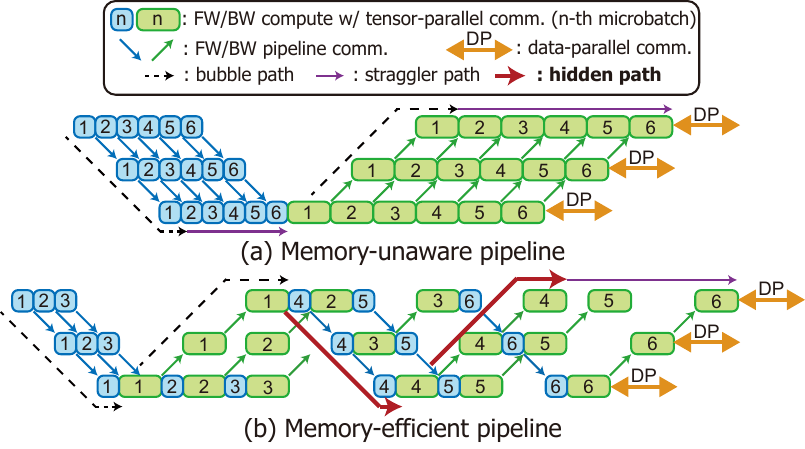}
  \caption{Pipeline scheduling.}
  \vspace{-5mm}
  \label{fig:background}
\end{figure}

Some researchers~\cite{piper, amp} have attempted to automatically tune the training configuration.
Typically, they have established a first-order pipeline latency model and performed an exhaustive search.
Let $C$ be the computational latency to process one microbatch, while $T_{com}^{(\cdot)}$ is the communication latency of each parallelization, $n\_mb$ is the number of microbatches, and $pp$ is the number of pipeline stages.
Then, the latency model from \cite{amp} is as follows:
\begin{equation}
\resizebox{0.8\columnwidth}{!}{%
$\begin{aligned}
     T_{prev} = (n\_mb-1)\cdot(C+T_{com}^{TP}) &+ pp \cdot (C+T_{com}^{TP}) \\
            + (pp-1) \cdot T_{com}^{PP} &+ T_{com}^{DP} \label{eq:prev}
\end{aligned}$
}
\end{equation}
The first term represents the runtime of stragglers in the pipeline ({\color{mypurple}purple arrows}).
The second and third terms capture the pipeline bubble computation and communication time (black dotted arrows), respectively.
The last term adds the data parallel communication time ({\color{myorange}orange arrow}).
Usually, profiled values are used for computation $C$, and document-specified bandwidth values are used for $T_{com}$.
However, this does not fit well with the modern memory-efficient schedules of \cref{fig:background}b, and does not consider the heterogeneous nature of real-world clusters.

\section{\thiswork Overview}
\label{sec:thiswork}

\begin{algorithm}
\caption{\thiswork Procedure}
\label{alg:procedure}
\begin{algorithmic}[1]
\footnotesize
 \renewcommand{\algorithmicrequire}{\textbf{Input:}}
 \REQUIRE $ $ \\
 $G$:\#GPUs,
 $bs_{global}$: global batch size, $M_{limit}$: memory limit per GPU \\
{\hspace{-6mm}\textbf{Output:}}\\
 $Conf$: parallelization configuration, \\$Map$: $\mathit{Conf} \rightarrow$ GPU mapping, $T$: execution latency\\
  \vspace{2mm}
  \STATE{$BW$\textleftarrow$\mathit{network\_profile()}$ \color{blue(ncs)}{ // Profile actual bandwidth matrix}}
  \STATE{$\mathit{Conf}_{best}$\textleftarrow$None$, $Map_{best}$\textleftarrow$None$, $T_{best}$\textleftarrow$\infty$} 
  \FOR {$\mathit{Conf}\in \{(pp, tp, dp)\ |\ pp\cdot tp\cdot dp=G, \quad pp, tp, dp \in \mathbb{N} \}$}
       \STATE{$bs_{mini} = bs_{global}/dp$ \color{blue(ncs)}{// Minibatch size}}
            \FOR{$bs_{micro} \in divisors$($bs_{mini}$)}
                \STATE{\color{blue(ncs)}{// Exclude OOM configurations (\cref{sec:mem_estimate})}}
                \STATE{\textbf{if}$\mathit{~MemEstimator(Conf, bs_{micro})} > M_{limit}$ \textbf{then} continue }
                \STATE{\color{blue(ncs)}{// \AutoConfig (\cref{sec:finegrained})}}   
                \WHILE{$Map \leftarrow \mathit{SA\_NextMap(Map)}$ }
                    \STATE{\color{blue(ncs)}{ // Estimate Latency (\cref{sec:latency})}}
                    \STATE{$T$\textleftarrow$\mathit{LatEstimator} (\mathit{Conf}, Map, bs_{mini}, bs_{micro}, BW)$} \\
                    \IF{{$T < T_{best}$}} 
                    \STATE{$Conf_{\mathit{best}}, Map_{\mathit{best}}, T_{\mathit{best}}$\textleftarrow$Conf, Map, T$}
                    \ENDIF
                \ENDWHILE
    \ENDFOR
  \ENDFOR
  \STATE{\textbf{return} $Conf_{\mathit{best}}, Map_{\mathit{best}}, T_{\mathit{best}}$}
 \end{algorithmic}
 \end{algorithm}

\cref{alg:procedure} shows the overall \thiswork procedure.
Similar to previous methods~\cite{amp, piper}, \thiswork finds the best configuration by examining the possible pipeline/tensor/data parallel combinations.
However, instead of relying on the document-specified link bandwidth, \thiswork considers the heterogeneous bandwidths of a real-world cluster by profiling them (line 1).
At the inner for loop, for each selected configuration ($Conf$) with each microbatch size ($bs_{micro}$), \thiswork checks the memory capacity constraint using \emph{\MemEstimate} (line 7, \cref{sec:mem_estimate}).
If it is runnable, \thiswork finds the best configuration for GPU mapping ($Map$) by \emph{\AutoConfig} with the simulated annealing (SA) algorithm (lines 9-15, \cref{sec:finegrained}).
SA keeps iterating the while loop for a specific time limit (e.g., 10s).
For each iteration, \emph{\LatencyEstimate} (denoted by $LatEstimator$) of \thiswork estimates the latency of the configuration of the current mapping (line 11, \cref{sec:latency}).
If the latency is the smallest one ever found, we store the current configuration/mapping/latency as the best case (line 13).
As a result, \thiswork returns the best configuration, mapping, and latency (line 18).

\section{\AutoConfig}
\label{sec:finegrained}

\begin{figure}[t]
\centering
 \includegraphics[width=\columnwidth]{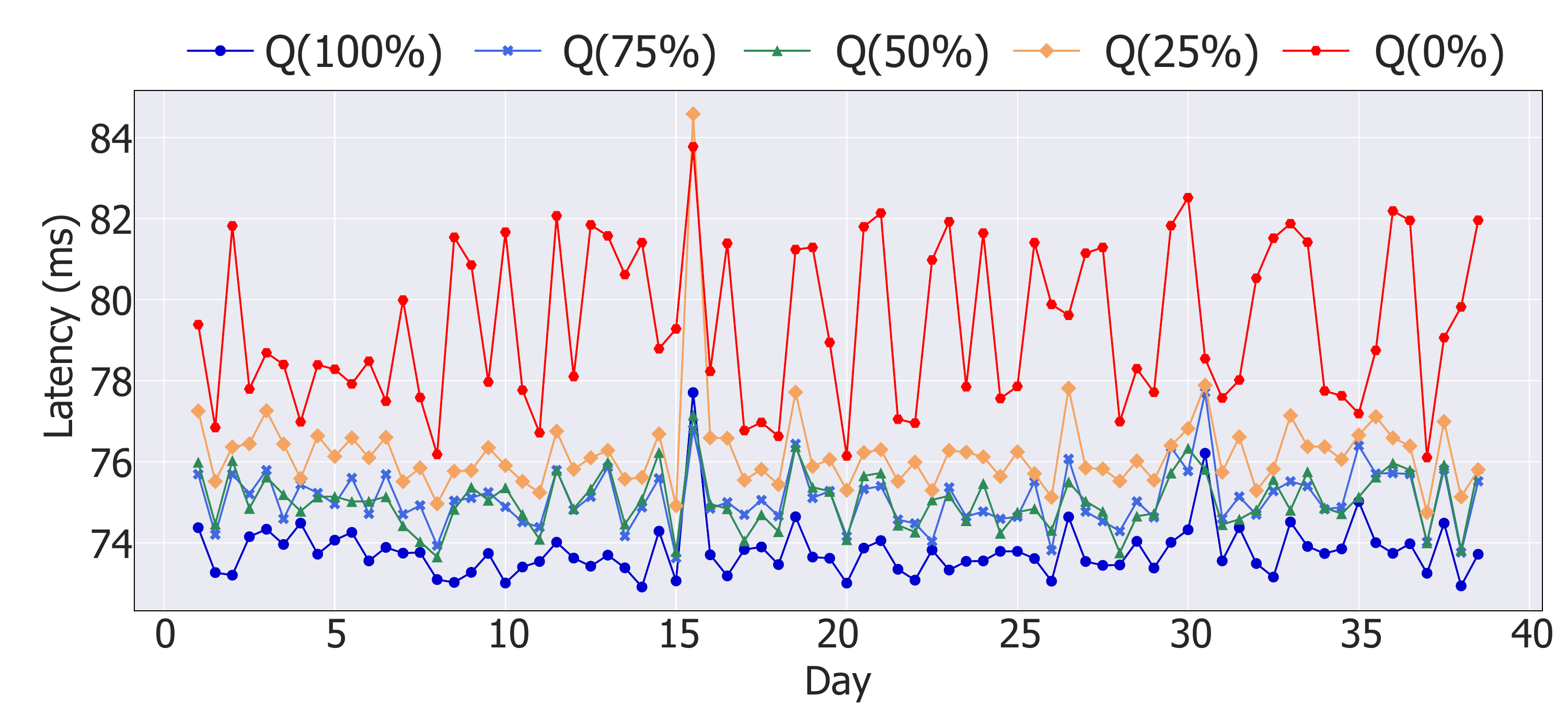} \vspace{-7mm}
 \caption{Inter-stage communication latency in a real-world cluster for 40 days.
 } \vspace{-5mm}
 \label{fig:hetero_issue}
\end{figure}

\begin{tcolorbox}[top=0mm, bottom=0mm]
\begin{itemize}[leftmargin=*]
\item \textbf{Key observation}: Real-world clusters have heterogeneous link bandwidths. 
\end{itemize}
\end{tcolorbox}

In \cref{fig:hetero_issue}, we provide a 40-day continuous profiling result of a commercial industry cluster (`high-end' environment from \S\ref{sec:eval:env}) to measure the actual peer-to-peer bandwidth of interconnect using mpiGraph~\cite{mpiGraph_github}.
Each line shows the latency corresponding to the order combination of 8 nodes.
We made sure that only the measurements were running to ensure that the measurements were isolated and reliable. 
From the plot, it is clear that the pairs of nodes exhibit different latencies even though they are designed to be equal.
Such phenomenon has been reported from multiple commercial clouds~\cite{plink, moody2009contention, vazhkudai2018design}, which indicates the practical need for considering the differences.

\begin{figure}[t]
\centering
 \includegraphics[width=\columnwidth]{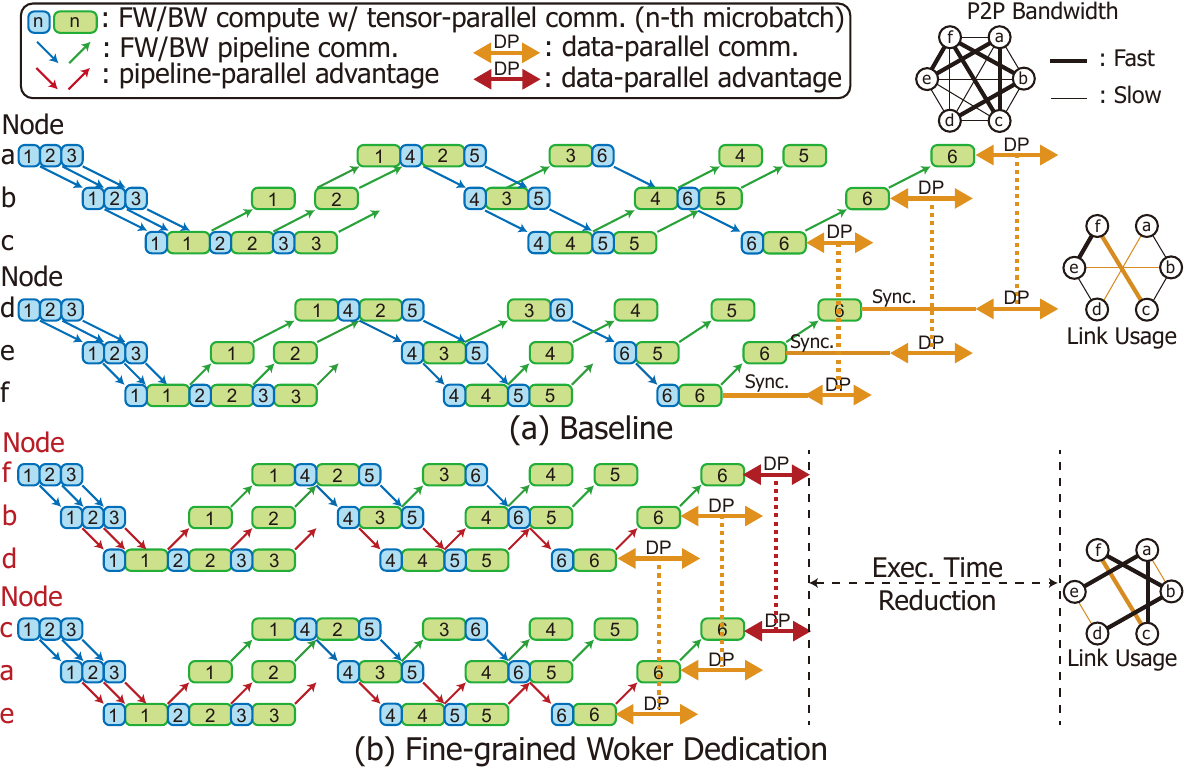} \vspace{-7mm}
 \caption{Baseline and \autoconfig schedule.
 } \vspace{-5mm}
 \label{fig:dedication}
\end{figure}

To embrace such heterogeneity in the parallel configuration, we advocate for dedicating individual logical workers to the physical GPUs in a fine-grained manner.
There are several rationales to this. 
First, the communications between the pipeline stages directly affect the execution time, but they take place only on a subset of the cluster network. 
By steering the traffic to travel through higher-speed links, the execution time can be reduced.
Furthermore, certain DP (i.e., all-reduce) traffic does not affect the execution time. 
As shown in \cref{fig:dedication}, only the DP communication of the earlier stages (a $\leftrightarrow$ d in (a) and f $\leftrightarrow$ c in (b)) are on the critical path.

In \cref{fig:dedication}, we provide a toy example of a six-node cluster based on the pipeline schedule example in \cref{fig:background}b of $pp=3$ and $dp=2$.
The bandwidth variance between the six servers is slightly exaggerated to have 2$\times$ difference between the slow links (thin lines) and fast links (bold lines).
As illustrated in \cref{fig:dedication}a, a naive configuration (alphabetical ordering) could result in a longer pipeline schedule.
The inter-stage communications between nodes (a $\rightarrow$ b, b $\rightarrow$ c, and d $\rightarrow$ e) are slower than other inter-stage communications.
The data parallel communication in the first stage (a $\leftrightarrow$ d) is much slower than in the last stage (c $\leftrightarrow$ f) due to the heterogeneity.

On the other hand, \cref{fig:dedication}b shows the pipeline from the \autoconfig.
We reordered/regrouped the nodes in the pipeline and changed the data parallel grouping.
For example, the pipeline group of node (a, b, c) is changed into node (f, b, d).
This minimizes the inter-stage communication
and reduces the data parallel communication on the critical path, as depicted with red arrows.

This problem essentially becomes finding a 1:1 mapping $f$ between logical workers and the GPUs.
Given a parallel configuration $pp, tp$, and $dp$, 
\begin{equation}
\resizebox{0.7\columnwidth}{!}{%
$\begin{aligned}
    f:W\rightarrow G&, \quad W =\mathbb{N}^{[1,pp]}\times\mathbb{N}^{[1,tp]}\times\mathbb{N}^{[1,dp]} \\
    &\forall w_i \neq w_j ,\ f(w_i) \neq f(w_j),
\end{aligned}$
}
\end{equation}
where $W$ is the set of logical workers and $G$ is the set of GPUs, where $|W|=|G|=pp\times tp\times dp$.
This problem is analogous to the classic multi-core mapping~\cite{murali2004bandwidth, hu2005energy}, and thus we use the popular simulated annealing (SA) to find $f$, similar to \cite{hu2005energy}.
We used three movements for the SA.
Regarding $f$ as a string, the first two movements are \emph{migration} (remove a single element to a random position), \emph{swap} (exchange two elements), and \emph{reverse} (take a substring and reverse its order).
Notably, the reverse move is based on our observation that the bidirectional bandwidths between a pair of nodes are often almost symmetric. 
The objective is to minimize the execution latency (\cref{sec:latency}).
We used 10 seconds for the SA time limit, and set $\alpha=.999$ for the temperature reduction coefficient. The overall time of the simulated annealing takes a few minutes.

\section{\LatencyEstimate}
\label{sec:latency}

\begin{tcolorbox}[top=0mm, bottom=0mm]
\begin{itemize}[leftmargin=*]
\item \textbf{Key observation}: Modern memory-efficient pipeline schedule contains hidden critical paths.
\end{itemize}
\end{tcolorbox}

A latency model is a necessary component to evaluate the mapping functions recommended by \autoconfig through SA. 
Actually, running the model in the target cluster consumes too much time and is not practical, especially in a shared cloud with long waiting queues. 

Existing approaches~\cite{amp, varuna} model the pipeline latency as \cref{eq:prev}, whose critical path comprises the bubble path and the straggler path.
However, we identify an additional hidden critical path that appears on the modern memory-efficient schedule, as depicted with {\color{red}red arrows} in \cref{fig:background}b.
As the key to the memory-efficient schedule is to interleave the forward and backward blocks (i.e., 1F1B), such hidden paths occur $(n\_mb/pp-1)$ times.
To reflect this, we divide the memory-efficient pipeline into bubbles ($T_{bubble}$), a straggler ($T_{straggler}$), and data parallel communication time ($T_{com}^{DP}$). 
With those, the total pipeline latency model of \thiswork ($T_{Pipette}$) is as follows:
\begin{equation}
\resizebox{0.75\columnwidth}{!}{%
$\begin{aligned}
     T_{Pipette} = T_{bubble}\cdot(n\_mb/pp) + T_{straggler} + T_{com}^{DP}, \label{eq:total}
\end{aligned}$
}
\end{equation}
where $T_{bubble}$ and $T_{straggler}$ can be calculated by following:
\begin{equation}
\resizebox{0.70\columnwidth}{!}{%
$\begin{aligned}
     &T_{bubble} = pp \cdot (C+T_{com}^{TP}) + (pp-1) \cdot T_{com}^{PP},\\
     &T_{straggler} = (pp-1) \cdot (C+T_{com}^{TP}). \label{eq:bubble_straggler}
\end{aligned}$
}
\end{equation}

For the individual terms, we model them as follows:
First, for the intra-node microbatch computation ($C$) and tensor parallel communication ($T^{TP}_{com}$), we use the profiled values as in previous works~\cite{varuna, amp}.
Optionally, we provide an extrapolated latency estimation model for other cluster sizes that have not been profiled, similar to our memory estimator (\cref{sec:mem_estimate}).

Second, for the pipeline parallel communication latency ($T_{com}^{PP}$), we define $B(g_1, g_2)$ as the pairwise bandwidth between $g_1$ and $g_2$ in $G$.
We set the pipeline parallel message size as $msg_{PP}$.
With a slight abuse of notation, we model $T^{PP}_{com}$ as:
\begin{equation}
\resizebox{0.65\columnwidth}{!}{%
$\begin{aligned}
      T_{com}^{PP} = \max_{y,z}\sum_{x=1}^{pp-1}\frac{2\cdot msg_{PP}}{B(G_{f(x, y, z)},G_{f(x+1, y, z)})}.
\end{aligned}$
}
\end{equation}
The $msg_{PP}$ is doubled to account for the forward and backward passes. 
$x,y$, and $z$ represent the ids of the pipeline-, tensor-, and data-parallel groups, respectively. 
Thus, the denominator represents the bandwidth between adjacent pipeline stages, and $T_{com}^{PP}$ takes the slowest of the end-to-end pipelines.

Third, for the data parallel communication term ($T_{com}^{DP})$, 
we assume the hierarchical-ring all-reduce algorithm, which contains two intra-node all-reduces and a single inter-node all-reduces.
We set the group of intra-node communicators (GPUs) on  pipeline stage $x$, tensor group $y$ as $W^{intra}_{x,y}$ and inter-node communicators (nodes) as $W^{inter}_{x,y}$.
Following the well-known all-reduce communication latency equation from~\cite{optimizationOfCollectiveComm}, the data parallel communication latency becomes:
\begin{equation}
\resizebox{.85\columnwidth}{!}{%
$\begin{aligned}
    T_{com}^{DP} = 
    \max_{y}\frac{4\cdot (|W^{intra}_{1, y}|-1) \cdot msg_{DP}}{(|W^{intra}_{1, y}|)\cdot{\min\limits_{w_1, w_2 \in W^{intra}_{1, y}}(B(G_{f(w_1)},G_{f(w_2)}))}}
  \\   +
    \max_{y}\frac{2\cdot (|W^{inter}_{1, y}|-1) \cdot msg_{DP}}{(|W^{inter}_{1, y}|)\cdot{\min\limits_{w_1, w_2 \in W^{inter}_{1, y}}(B(G_{f(w_1)},G_{f(w_2)}))}}.
\end{aligned}$
}
\end{equation}
Only the DP communication of stage 1 on the critical path is considered, and the all-reduce delay depends on the slowest link in participation.
To verify the effect of the new latency model and bandwidth profiling, we compared the estimation results of \latencyestimate with \cite{amp} in \cref{fig:mem_issue}a.
\cite{amp} shows $23.18\%$ mean absolute percentage error (MAPE), whereas \thiswork shows $5.87\%$.
\cite{amp} fails to estimate latencies accurately because it does not consider the memory-efficient schedule and uses the ideal bandwidth in the communication-related terms.

\section{\MemEstimate}
\label{sec:mem_estimate}

\begin{figure}[t]
\centering
 \includegraphics[width=.95\columnwidth]{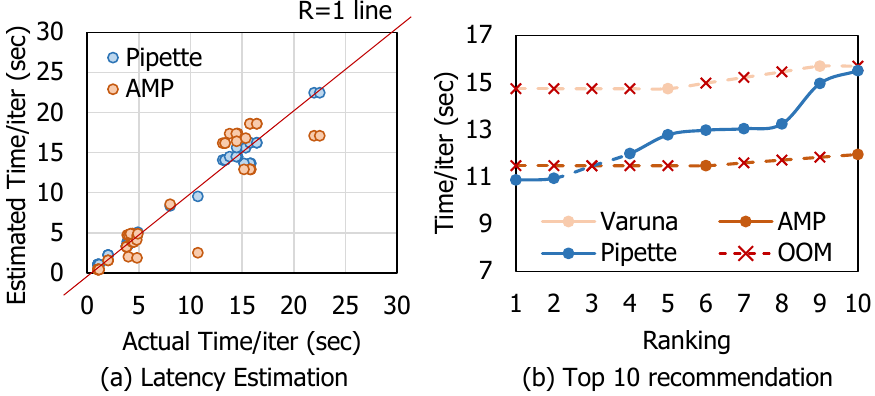} \vspace{-3mm}
 \caption{Latency estimation and top 10 recommendation of baselines and \thiswork.
 } \vspace{-5mm}
 \label{fig:mem_issue}
\end{figure}

\begin{tcolorbox}[top=0mm, bottom=0mm]
\begin{itemize}[leftmargin=*]
\item \textbf{Key observation}: Prior art often recommends OOM solutions, hindering them from being practical.
\end{itemize}
\end{tcolorbox}

When the existing approaches recommend configurations, they fail to consider the memory requirements. To demonstrate this, we plot the top 10 configurations from AMP~\cite{amp} and Varuna~\cite{varuna} in 
\cref{fig:mem_issue}b.
The experiment was conducted on the mid-range cluster in \cref{sec:eval:env}.
Eight of the top 10 suggested configurations incur OOM in both methods, including the top recommendations.
This significantly harms the practicality because users need to try the solutions one by one until they find a feasible configuration.
Therefore, when searching for configurations, we integrate an accurate \memestimate.
A common way to estimate the memory requirement is by dividing the model size by the number of stages and tensor-parallel ways and then approximating the activation size by considering the layer structures~\cite{heuristic_mem_model}.

However, with pipelining, such a simple model significantly underestimates the memory usage, especially the auxiliary structures of the training framework and external libraries~\cite{ms_automem_estimate}.
Taking advantage of the fact that LLMs are composed of the same transformer blocks, 
we model the memory requirement as a function of several system and model parameters.
When we approximate them with a multi-layer perceptron (MLP): 
\begin{equation}
\resizebox{0.8\columnwidth}{!}{%
$\begin{aligned}
     M_{max} = MLP(n\_gpus, n\_layers, n\_hiddens, n\_heads,\\ tp, pp, dp, bs_{micro}, bs_{mini}, bs_{global}).
\end{aligned}$
}
\end{equation}

The inputs include the configuration parameters (\cref{alg:procedure}) in addition to the number of GPUs ($n\_gpus$), layers ($n\_layers$), hidden dimensions ($n\_hidden$), and attention heads ($n\_heads$). 
To train the model, we use the profiled data from all possible configurations using up to four cluster nodes (32 GPUs) and validate the model extrapolation up to 128 GPUs. 
The MLP model has five layers with 200 hidden sizes and is trained for 50,000 iterations with the profiled data.
This training process is required for each cluster only once because the trained model can be used afterward for all other configurations.
When \memestimate decides whether a configuration is runnable, it sets a soft margin to stably recommend runnable configurations.
As a result, \thiswork recommends more runnable results than baselines, as shown in \cref{fig:mem_issue}b.

\section{Evaluation}
\label{sec:eval}

\subsection{Environments}
\label{sec:eval:env}
\begin{table}
\scriptsize
\centering
\caption{Experimental Environment}\label{tab:environment}
\begin{tabular}{ccc}
\toprule
 \multirowcell{5}{Mid-range\\Cluster\\(16 Nodes)} 
 & GPU & 8$\times$ NVIDIA V100 \\
& CPU & 2$\times$ Xeon Gold 6142, 16 cores\\
& Memory & 768GB DDR4 ECC \\
& Inter-node & Infiniband EDR (100Gbps)\\
& Intra-node & NVLink (300GBps) \\

\cmidrule(lr){2-3}

 \multirowcell{5}{High-end\\Cluster\\(16 Nodes)} 
 & GPU & 8$\times$ NVIDIA A100 \\
& CPU & 2$\times$ EPYC 7543, 32 cores\\
& Memory & 1TB DDR4 ECC \\
& Inter-node & Infiniband HDR (200Gbps)\\
& Intra-node & NVLink (600GBps) \\

 \bottomrule
\end{tabular}
\vspace{-3mm}
\end{table} 

\textbf{Clusters.} \cref{tab:environment} shows the experimental environment we used for evaluation.
We used two clusters:
The V100 (`Mid-range') cluster has an inter-node connection with Infiniband EDR (100Gbps) and an intra-node connection with NVLink (300GBps).
The A100 (`High-end') cluster has an inter-node connection with Infiniband HDR (200Gbps) and an intra-node connection with NVSwitch (600GBps).
We set a cluster with 128 GPUs (16 nodes) as default in both cases. 

\textbf{Framework and models.} We used the publicly available source code of Megatron-LM~\cite{megatrongithub} as our baseline LLM training framework.
For interconnect profiling, we ran NCCL-tests~\cite{nccl_test_github}.
We tested GPT models of sizes up to 3.1B and 11.1B parameters, which reach the GPU memory limit in the mid-range and high-end clusters, respectively.
We evaluated the total minibatch size from 128 to 512 and the microbatch size from 1 to 8 for experiments and adopted other hyperparameters from~\cite{megatron-lm}. 

\textbf{Baselines.} We chose manually tuned method Megatron-LM (MLM)~\cite{megatron-lm}, Varuna (VR)~\cite{varuna} and AMP~\cite{amp} as baselines.
Megatron-LM generally tunes the number of GPUs per node as a tensor parallel way ($tp=8$).
Therefore, if not stated, we fixed $tp=8$ in the MLM baseline.
Varuna emphasizes using the pipeline parallel-only configuration for LLM training and provides successful speedup over~\cite{sc21megatron}.
We use the default configuration suggested by its open-sourced code.
AMP is the state-of-the-art automatic configurator for 3D parallelism, so we tested it as a baseline.
Because AMP often recommends out-of-memory cases, we manually tested them one by one from the top recommendation until we reached a runnable configuration.

\subsection{Speedup and Ablation}

\begin{figure}[t]
\centering
 \includegraphics[width=\columnwidth]{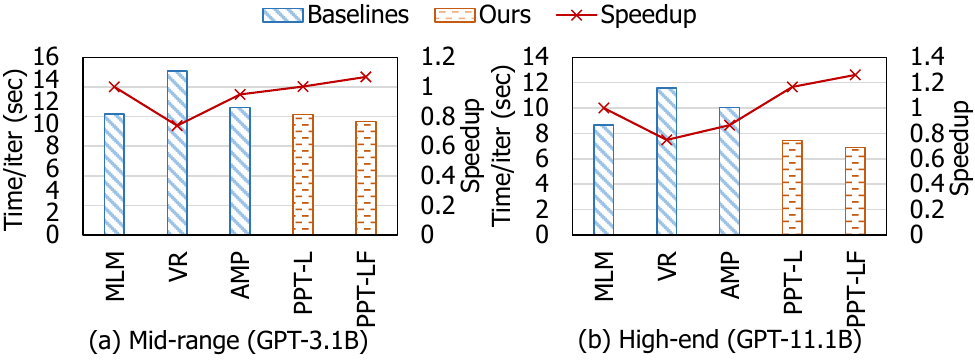} \vspace{-3mm}
 \caption{Training time and speedup of \thiswork and the baselines.
 }\vspace{-5mm}
 \label{fig:speedup}
\end{figure}

\cref{fig:speedup} shows the training time and speedup of \thiswork compared to the baselines.
MLM provides better training throughput than other baselines because it is tuned for the memory-efficient pipeline, while other baselines recommend configurations based on the memory-unaware one.
Therefore, we normalized the speedup to MLM.
We also give ablation in \cref{fig:speedup} to illustrate the effect of each scheme.
`PPT-L' means \thiswork only with the elaborately designed \latencyestimate with \memestimate applied.
`PPT-LF' denotes \thiswork with both \latencyestimate and \autoconfig.

In mid-range and high-end clusters, PPT-L provides 1.36$\times$, 1.56$\times$ speedup over Varuna (VR).
VR uses a configuration without tensor parallelism.
However, in both our configurations, VR induces too much network overhead, especially due to the existence of the hidden critical path (\cref{sec:latency}).
AMP is much faster than VR, but PPT-L still shows 1.06$\times$ and 1.35$\times$ speedup over it.
This comes from the ability of \thiswork to estimate the execution latency correctly.
\thiswork with \autoconfig (PPT-LF) further exploits the heterogeneity of a real-world cluster, showing 1.12$\times$, 1.46$\times$ speedup over AMP in mid-range and high-end clusters, respectively.
As a result, \thiswork provides 1.07$\times$ and 1.26$\times$ speedup over the manually tuned baseline (MLM), which requires manual trials to find a fast configuration.
Additionally, \thiswork tends to show further speedup on the high-end cluster, which handles larger models.

\subsection{Analysis of \MemEstimate}
\label{sec:eval:memory}

\begin{figure}[t]
\centering
 \includegraphics[width=.95\columnwidth]{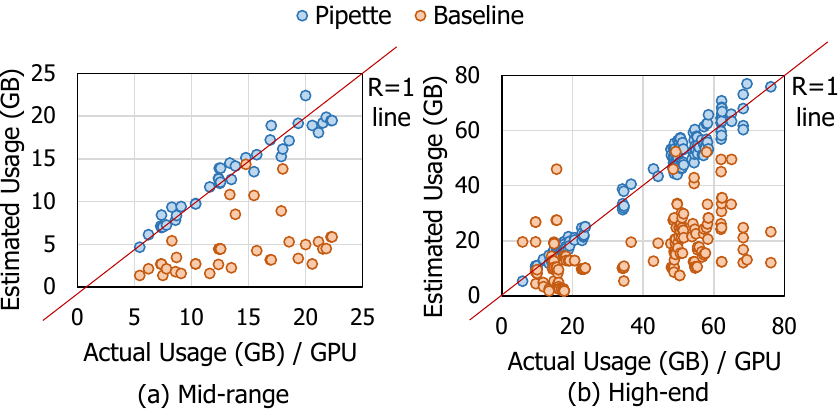} \vspace{-3mm}
 \caption{Memory estimation results of \thiswork and the baseline.
 } \vspace{-5mm}
 \label{fig:memory}
\end{figure}

When searching for the best configuration, \thiswork checks whether the configuration does not exceed the GPU memory limit.
To check the practicality of \memestimate, we compare it with~\cite{heuristic_mem_model}.
\cref{fig:memory} plots how well \thiswork and the baseline estimate the actual maximum memory usage.
We collected 215 data points, representing the estimated and actual GPU memory usage with various model and parallel configurations.
The baseline underestimates the maximum memory usage because it does not consider the usage from the training framework and external libraries~\cite{ms_automem_estimate}.
Therefore, the baseline provides the mean absolute percentage error (MAPE) of $65.71\%$ and $59.49\%$, while \Memestimate of \thiswork only shows $7.39\%$ and $6.42\%$ in mid-range and high-end clusters, respectively.
This demonstrates the importance of a dedicated \memestimate for an automatic configurator.

\subsection{Analysis of Configuration Overhead}

\begin{table}
\scriptsize
\centering
\caption{Configuration Overhead of \thiswork}\label{tab:overhead}
\begin{tabular}{ccccc}
\toprule
Cluster & \multicolumn{2}{c}{Mid-range} & \multicolumn{2}{c}{High-end} \\
\cmidrule(lr){2-3}
\cmidrule(lr){4-5}
\#Nodes (Model) & 8 (1.1B) & 16 (3.1B) & 8 (8.1B) & 16 (11.1B) \\
\midrule
Bandwidth Profiling & 58.13 sec. & 119.62 sec. & 113.67 sec. & 239.21 sec. \\
Simulated Annealing & 640.38 sec. & 790.51 sec. & 640.23 sec. & 769.91 sec. \\
Memory Estimation & 0.03 sec. & 0.04 sec. & 0.03 sec. & 0.05 sec. \\
\cmidrule(lr){1-5}
Total Conf. Time & 10.71 min. & 13.23 min. & 10.71 min. & 16.85 min. \\
Overhead  & 0.03\% & 0.03\% & 0.02\% & 0.05\% \\

\midrule
AMP (300K) & 30.10 days & 37.74 days & 36.86 days & 34.85 days \\
\thiswork (300K)  & 29.13 days & 35.42 days & 31.61 days & 23.89 days  \\
Time Saving & 0.97 days & 2.33 days & 5.25 days & 10.97 days \\

 \bottomrule
\end{tabular}
\end{table} 

Compared to other works, \thiswork introduces bandwidth profiling, simulated annealing-based worker mapping, and \memestimate.
Therefore, analyzing the overhead of \thiswork is meaningful, while it is negligible compared to LLM training time.
In \cref{tab:overhead}, we demonstrate the configuration overhead of \thiswork.
The simulated annealing takes most of the total overhead.
The total number of nodes is the main factor in determining the problem size of the simulated annealing algorithm.
Therefore, the total time is similar in both mid-range and high-end clusters when the number of nodes is the same.
The overhead means the portion of configuration time when training the full iteration (300K) following~\cite{megatron-lm}.
The overhead is less than 0.05\% of the total training time, and the benefit is 0.97-10.97 days over the configuration recommended by AMP.

\subsection{Sensitivity Studies}

\begin{figure}[t]
\centering
 \includegraphics[width=\columnwidth]{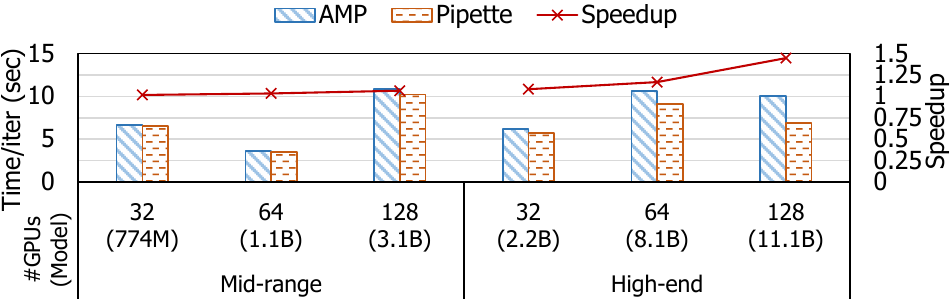} \vspace{-6mm}
 \caption{Cluster and model size scalability of \thiswork.
 } \vspace{-5mm}
 \label{fig:cluster_sensi}
\end{figure}

\textbf{Cluster/Model size.}
While we use a cluster with 128GPUs as the default, it is essential to check \thiswork still provides speedup with different numbers of GPUs (from 32 to 128).
\cref{fig:cluster_sensi} shows the training time and speedup of \thiswork over AMP with various numbers of GPUs.
We weak-scaled the model size with the number of GPUs, following~\cite{megatron-lm, zero}.
In clusters with smaller numbers of GPUs than the default setting, the heterogeneity appears less, so \thiswork shows a smaller speedup but still brings 1.02-1.17$\times$ speedup.

\begin{figure}[t]
\centering
 \includegraphics[width=\columnwidth]{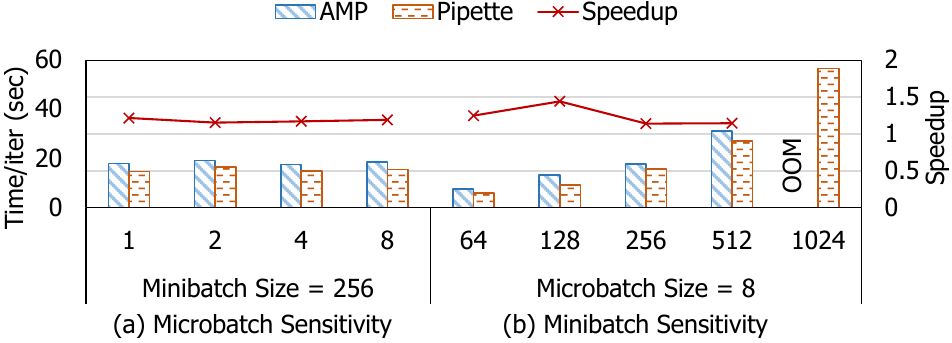} \vspace{-7mm}
 \caption{Microbatch and minibatch size sensitivity of \thiswork.
 } \vspace{-5mm}
 \label{fig:batch_sensi}
\end{figure}
\textbf{Micro/Minibatch size.}
Recent works use microbatch sizes from 1 to 8 and minibatch sizes from 64 to 1K, so we checked the micro/minibatch size sensitivity of \thiswork, in \cref{fig:batch_sensi}.
For the experiment, we ran \thiswork and AMP to find configurations when the micro/batch size is fixed.
For the microbatch size sensitivity, we fixed minibatch size to 256.
In minibatch size sensitivity, we used microbatch size with 8.
In all settings, \thiswork provides stable 1.14-1.44$\times$ speedup over AMP, which shows the practicality of \thiswork.

\section{Conclusion}
We propose \thiswork, an automatic fine-grained LLM training configurator for real-world clusters.
\thiswork provides significant speedup over the baselines by devising an accurate, critical path-based latency model and dedicating each worker in a fine-grained manner.
In addition, to the best of our knowledge, \thiswork is only the configurator that recommends configurations meeting the memory capacity constraints. 
This greatly enhances the practicality of \thiswork.

\section*{Acknowledgement}
This work was supported by 
Samsung Electronics Co., Ltd (IO221111-03540-01),
and the National Research Foundation of Korea (NRF) grants (2022R1C1C1011307).

\defbibenvironment{bibliography}
  {\trivlist}
  {\endtrivlist}
  {\item
   \printtext[labelnumberwidth]{%
     \printfield{labelprefix}%
      \printfield{labelnumber}}%
   \addspace}

\begin{footnotesize}
\printbibliography

@inproceedings{
amp,
author = {Li, Dacheng and Wang, Hongyi and Xing, Eric and Zhang, Hao},
title = {{AMP: Automatically Finding Model Parallel Strategies with Heterogeneity Awareness}},
booktitle = {NeurIPS}, 
year = {2022}
}

@inproceedings{
piper,
title={{Piper: Multidimensional Planner for DNN Parallelization}},
author={Jakub Tarnawski and Deepak Narayanan and Amar Phanishayee},
booktitle={NeurIPS},
year={2021}
}

@article{bert,
  title={{Bert: Pre-training of Deep Bidirectional Transformers for Language Understanding}},
  author={Devlin, Jacob and Chang, Ming-Wei and Lee, Kenton and Toutanova, Kristina},
  journal={arXiv preprint arXiv:1810.04805},
  year={2018}
}

@article{gpt2,
  title={{Language Models Are Unsupervised Multitask Learners}},
  author={Radford, Alec and Wu, Jeffrey and Child, Rewon and Luan, David and Amodei, Dario and Sutskever, Ilya and others},
  journal={OpenAI blog},
  year={2019}
}

@inproceedings{gpt3,
 author = {Brown, Tom and Mann, Benjamin and Ryder, Nick and Subbiah, Melanie and Kaplan, Jared D and Dhariwal, Prafulla and Neelakantan, Arvind and Shyam, Pranav and Sastry, Girish and Askell, Amanda and Agarwal, Sandhini and Herbert-Voss, Ariel and Krueger, Gretchen and Henighan, Tom and Child, Rewon and Ramesh, Aditya and Ziegler, Daniel and Wu, Jeffrey and Winter, Clemens and Hesse, Chris and Chen, Mark and Sigler, Eric and Litwin, Mateusz and Gray, Scott and Chess, Benjamin and Clark, Jack and Berner, Christopher and McCandlish, Sam and Radford, Alec and Sutskever, Ilya and Amodei, Dario},
 booktitle = {NeurIPS},
 title = {{Language Models are Few-Shot Learners}},
 year = {2020}
}

@article{megatron-lm,
  title={{Megatron-LM: Training Multi-billion Parameter Language Models Using Model Parallelism}},
  author={Shoeybi, Mohammad and Patwary, Mostofa and Puri, Raul and LeGresley, Patrick and Casper, Jared and Catanzaro, Bryan},
  journal={arXiv preprint arXiv:1909.08053},
  year={2019}
}

@inproceedings{sc21megatron,
author = {Narayanan, Deepak and Shoeybi, Mohammad and Casper, Jared and LeGresley, Patrick and Patwary, Mostofa and Korthikanti, Vijay and Vainbrand, Dmitri and Kashinkunti, Prethvi and Bernauer, Julie and Catanzaro, Bryan and Phanishayee, Amar and Zaharia, Matei},
title = {{Efficient Large-Scale Language Model Training on GPU Clusters Using Megatron-LM}},
year = {2021},
booktitle = {SC}
}

@INPROCEEDINGS{zero,
  author={Rajbhandari, Samyam and Rasley, Jeff and Ruwase, Olatunji and He, Yuxiong},
  booktitle={SC}, 
  title={{ZeRO: Memory Optimizations Toward Training Trillion Parameter Models}}, 
  year={2020}
}

@inproceedings{pipedream,
author = {Narayanan, Deepak and Harlap, Aaron and Phanishayee, Amar and Seshadri, Vivek and Devanur, Nikhil R. and Ganger, Gregory R. and Gibbons, Phillip B. and Zaharia, Matei},
title = {{PipeDream: Generalized Pipeline Parallelism for DNN Training}},
year = {2019},
booktitle = {SOSP}
}

@inproceedings{optimuscc,
author = {Song, Jaeyong and Yim, Jinkyu and Jung, Jaewon and Jang, Hongsun and Kim, Hyung-Jin and Kim, Youngsok and Lee, Jinho},
title = {{Optimus-CC}: Efficient Large {NLP} Model Training with {3D} Parallelism Aware Communication Compression},
year = {2023},
booktitle = {ASPLOS},
}

@misc{nccl_test_github,
title={{NVIDIA NCCL Tests}},
howpublished = {https://github.com/NVIDIA/nccl-tests},
}

@misc{mpiGraph_github,
title={{LLNL mpiGraph Tests}},
howpublished = {https://github.com/LLNL/mpiGraph},
}

@software{megatrongithub,
  author       = {Jared Casper and
                  Mostofa Patwary and
                  Boris Fomitchev and
                  Evelina and
                  lmcafee-nvidia and
                  Raul Puri and
                  Nako Sung and
                  Stas Bekman and
                  Akhilesh Gotmare and
                  David E. Weekly and
                  Deepak Narayanan and
                  Devrim and
                  Heungsub Lee and
                  Kazuhiro Yamasaki},
  title        = {NVIDIA/Megatron-LM: v2.5},
  month        = aug,
  year         = 2021,
  publisher    = {Zenodo},
  version      = {v2.5},
  doi          = {10.5281/zenodo.5181820}
}

@inproceedings{varuna,
author = {Athlur, Sanjith and Saran, Nitika and Sivathanu, Muthian and Ramjee, Ramachandran and Kwatra, Nipun},
title = {{Varuna: Scalable, Low-Cost Training of Massive Deep Learning Models}},
year = {2022},
booktitle={EuroSys}
}

@article{optimizationOfCollectiveComm,
author = {Rajeev Thakur and Rolf Rabenseifner and William Gropp},
title ={{Optimization of Collective Communication Operations in MPICH}},
journal = {IJHPCA},
year = {2005},
pages = {49-66}
}

@inproceedings{plink,
 author = {Luo, Liang and West, Peter and Nelson, Jacob and Krishnamurthy, Arvind and Ceze, Luis},
 booktitle = {MLSys},
 title = {{PLink: Discovering and Exploiting Locality for Accelerated Distributed Training on the public Cloud}},
 year = {2020}
}

@article{hu2005energy,
  title={{Energy-and performance-aware mapping for regular NoC architectures}},
  author={Hu, Jingcao and Marculescu, Radu},
  journal={IEEE TCAD},
  year={2005}
}

@inproceedings{murali2004bandwidth,
  title={{Bandwidth-constrained mapping of cores onto NoC architectures}},
  author={Murali, Srinivasan and De Micheli, Giovanni},
  booktitle={DATE},
  year={2004}
}

@misc{heuristic_mem_model,
author={Bricken, Trenton},
title={{Transformer Memory Requirements}},
note = {\url{https://www.trentonbricken.com/TransformerMemoryRequirements/}},
year={2022}
}

@techreport{moody2009contention,
  title={Contention-free routing for shift-based communication in MPI applications on large-scale InfiniBand clusters},
  author={Moody, Adam},
  year={2009},
  institution={Lawrence Livermore National Lab.(LLNL), Livermore, CA (United States)}
}

@inproceedings{vazhkudai2018design,
  title={The design, deployment, and evaluation of the CORAL pre-exascale systems},
  author={Vazhkudai, Sudharshan S and De Supinski, Bronis R and Bland, Arthur S and Geist, Al and Sexton, James and Kahle, Jim and Zimmer, Christopher J and Atchley, Scott and Oral, Sarp and Maxwell, Don E and others},
  booktitle={SC},
  year={2018}
}

@inproceedings{ms_automem_estimate,
  title={{Estimating GPU Memory Consumption of Deep Learning Models}},
  author={Gao, Yanjie and Liu, Yu and Zhang, Hongyu and Li, Zhengxian and Zhu, Yonghao and Lin, Haoxiang and Yang, Mao},
  booktitle={ESEC/FSE},
  year={2020}
}
\end{footnotesize}

\end{document}